# The Fluctuation Theorem and Lyapunov Weights

Owen Jepps[†]

School of Science

Griffith University

Brisbane, Qld 4111 Australia

Denis J. Evans

Research School of Chemistry

Australian National University

Canberra, ACT 0200 Australia

and

Debra J. Searles

School of Science

Griffith University

Brisbane, Qld 4111 Australia

[†]current address, Department of Chemical Engineering, University of Queensland, Brisbane, Qld 4072 Australia

**Abstract**

The Fluctuation Theorem (FT) is a generalisation of the Second Law of Thermodynamics that applies to small systems observed for short times. For thermostatted systems it gives the probability ratio that entropy will be consumed rather than produced. In this paper we derive the Transient and Steady State Fluctuation Theorems using Lyapunov weights rather than the usual Gibbs weights. At long times the Fluctuation Theorems so derived are identical to those derived using the more standard Gibbs weights.

# 1. INTRODUCTION

The fluctuation theorem [1-10] (FT) gives a formula for the logarithm of the probability ratio that the time averaged dissipative flux $\bar{J}_t$, takes a value, A, to minus the value, -A. This formula is an analytic expression that gives the probability, for a finite system and for a finite time, t, that the dissipative flux flows in the reverse direction to that required by the Second Law of Thermodynamics. The FT is consistent with the Second Law in that as either the number of particles (N) or the averaging time becomes infinite, the probability of Second Law violations goes to zero. The Fluctuation Theorem can therefore be considered as a generalisation of the Second Law of Thermodynamics for finite systems and finite times. For steady state trajectory segments of duration t, (*i.e.* those trajectory segments that are initiated long after the application of a field so that the system has reached a steady state), the Steady State FT (SSFT) is only true in the long time limit. Evans and Searles [2-7] have shown that if transient trajectories are considered rather than steady state trajectories, a Transient FT (TFT) that is true at all times can be derived. In the transient experiment the value of $\bar{J}_t$ is obtained by averaging along a trajectory segment that is initially (at t = 0) sampled from a known initial distribution (such as an equilibrium distribution), but to which a field is subsequently applied (t ≥ 0). When the nonequilibrium steady state is unique, one would expect the asymptotic convergence of the TFT to the SSFT since averages over transient segments should then approach (to within arbitrary accuracy as t → ∞) those taken over nonequilibrium steady state segments. The TFT for thermostatted nonequilibrium systems has recently been verified experimentally [10]. For a pedagogical review on the FT, see reference [9].

The TFT considers the response of an ensemble of systems to an applied dissipative field. If the system is thermostatted, it may reach a steady state after a Maxwell time, $\tau_M$. In the original derivation [2], it was supposed that the initial ensemble was the microcanonical ensemble, the dynamics was isoenergetic and the Liouville measures were used. The microcanonical initial ensemble was chosen because the probability of observing trajectories originating in a specified phase volume is simply proportional to the measure of that volume. However, it is straightforward to apply the same procedure to an arbitrary initial ensemble, and an analytical form of the TFT that is valid at all times and for a range of ensembles and dynamics has been obtained [5, 9].

The original heuristic derivation of the SSFT [1] employed dynamical weights for computing the probabilities of occurrence of trajectory segments. Such weights are completely different to the Boltzmann and Gibbs (static) weights that are so familiar in equilibrium statistical mechanics. In ECM2 [1], it was argued that if one trajectory segment is much more *unstable* than a second trajectory segment (*i.e.* the first has a much larger sum of positive Lyapunov exponents than the second segment), then there should be much *less* probability of observing other trajectory segments near the first segment than near the second segment.

Later Gallavotti and Cohen (GC) [8] pointed out that these dynamical weights were already known in dynamical systems theory. In 1995 they gave a more rigourous derivation of the SSFT, based on Markov partitioning of the steady state attractor and employing the so-called chaotic hypothesis, [8]. The systems considered were restricted to those undergoing isoenergetic dynamics.

Here we extend these arguments so we can employ a partitioning of dynamical weights for a range of different ensembles and dynamics. Unlike the work of Gallavotti and Cohen, the derivation is developed by first considering transient trajectories evolving from an initial equilibrium distribution. This considerably simplifies the derivation because the space over which the partition is required is not a fractal object. We show with reasonable rigour, that dynamical weights based on exponentials of sums of positive, finite time, Lyapunov exponents lead to TFTs that are identical to those derived previously using static weights, for various forms of initial ensemble (canonical, microcanonical etc.) and various forms of thermostatting (constant temperature, constant energy etc.).

## 2. A TRANSIENT FLUCTUATION THEOREM FROM LYAPUNOV MEASURES

The TFT was derived [2, 5, 9] using Liouville measures, whereas the ECM2 and GC derivations of the SSFT utilised Lyapunov weights. Here we consider a derivation of the TFT from Lyapunov weights.

Consider an N-particle system with coordinates and peculiar momenta, $\{\mathbf{q}_1, \mathbf{q}_2, ..\mathbf{q}_N, \mathbf{p}_1, ..\mathbf{p}_N\} \equiv (\mathbf{q}, \mathbf{p}) \equiv \mathbf{\Gamma}$. The internal energy of the system is $H_0 \equiv \sum_{i=1}^{N} p_i^2 / 2m + \Phi(\mathbf{q}) = K + \Phi$ where $\Phi(\mathbf{q})$ is the interparticle potential energy, which is a function of the coordinates of all the particles, $\mathbf{q}$ and K is the total peculiar kinetic energy. In the presence of an external field $F_e$, the thermostatted equations of motion are taken to be [11],

$$\dot{\mathbf{q}}_i = \mathbf{p}_i / m + \mathbf{C}_i(\mathbf{\Gamma}) \bullet \mathbf{F}_e$$
$$\dot{\mathbf{p}}_i = \mathbf{F}_i(\mathbf{q}) + \mathbf{D}_i(\mathbf{\Gamma}) \bullet \mathbf{F}_e - \alpha(\mathbf{\Gamma})\mathbf{p}_i \tag{1}$$

where $\mathbf{F}_i(\mathbf{q}) = -\partial\Phi(\mathbf{q})/\partial\mathbf{q}_i$, $\alpha$ is the thermostat multiplier which in this case is applied to the peculiar momenta, and $\mathbf{C}_i$ and $\mathbf{D}_i$ couple the system to the external field, $\mathbf{F}_e$. The dissipative flux is given by $\mathbf{J}$ where $\dot{H}_0^{ad} \equiv -\mathbf{J}(\mathbf{\Gamma})V \bullet \mathbf{F}_e$, where the superscript 'ad' says that the derivative should be carried out adiabatically (*i.e.* in the absence of the thermostat).

Suppose the autonomous equations of motion (1), are written

$$\dot{\mathbf{\Gamma}} = \mathbf{G}(\mathbf{\Gamma}). \tag{2}$$

It is trivial to see that the equation of motion for an infinitesimal phase space tangent vector, $d\mathbf{\Gamma}$, can be written as:

$$d\dot{\mathbf{\Gamma}} = \mathbf{T}(\mathbf{\Gamma}) \bullet d\mathbf{\Gamma}, \tag{3}$$

where $\mathbf{T} \equiv \partial\mathbf{G}(\mathbf{\Gamma})/\partial\mathbf{\Gamma}$ is the stability matrix for the flow. The propagation of the tangent vectors is therefore given by,

$$d\mathbf{\Gamma}[t;\mathbf{\Gamma}(0),d\mathbf{\Gamma}(0)] = \mathbf{L}[t;\mathbf{\Gamma}(0)] \bullet d\mathbf{\Gamma}(0) \tag{4}$$

where the propagator is:

$$\mathbf{L}[t;\mathbf{\Gamma}(0)] = \exp_L\left(\int_0^t ds \mathbf{T}(\mathbf{\Gamma}(s))\right), \tag{5}$$

and $\exp_L$ is a left time-ordered exponential. We note that $d\mathbf{\Gamma}[t;\mathbf{\Gamma}(0),d\mathbf{\Gamma}(0)]$ is a function of both the initial phase, $\mathbf{\Gamma}(0)$, and the initial tangent vector, $d\mathbf{\Gamma}(0)$.

The long time evolution of these tangent vectors are used to determine the Lyapunov spectrum for the system. The Lyapunov exponents represent the asymptotic rates of divergence of nearby points in phase space. The Lyapunov exponents are defined as:

$$\{\lambda_i[\mathbf{\Gamma}(0)]; i = 1,..2dN\} = \lim_{t \to \infty} \frac{1}{2t} \ln\left(\text{eigenvalues}\left(\mathbf{L}[t;\mathbf{\Gamma}(0)]^T \bullet \mathbf{L}[t;\mathbf{\Gamma}(0)]\right)\right), \tag{6}$$

It is frequently assumed that the system is ergodic (i.e. that time and ensemble averages are equivalent) and thus in the long time limit these eigenvalues are independent of the initial phase, $\mathbf{\Gamma}(0)$. By convention the exponents are ordered such that $\lambda_1 > \lambda_2 > \ldots > \lambda_{2d_cN}$, where $d_C$ is the Cartesian dimension of the system. A system is said to be chaotic if at least one exponent is positive (*i.e.* $\lambda_1 > 0$).

Consider an ensemble of systems which is initially characterised by a distribution $f(\mathbf{\Gamma},0)$. For simplicity, we assume that the initial distribution is symmetric under the time reversal mapping, i.e. $f(\mathbf{q},\mathbf{p},0) = f(\mathbf{q},-\mathbf{p},0)$[1]. Suppose that a phase space trajectory evolves from $\mathbf{\Gamma}_0(0)$ at t = 0 to $\mathbf{\Gamma}_0(t)$ at time t. We call this trajectory the mother trajectory. We also consider the evolution of a set of neighbouring phase points, $\mathbf{\Gamma}(0)$, that begin at time zero, within some fixed region of size determined by $d\Gamma$: $0 < \Gamma_\alpha(0) - \Gamma_{0,\alpha}(0) = \delta\Gamma_\alpha(0) < d\Gamma$, $\forall \alpha = 1,..2d_CN$ ($\Gamma_\alpha$ is the $\alpha$th component of the phase space vector $\mathbf{\Gamma}$, and $\Gamma_{0,\alpha}$ is the $\alpha$th component of the vector $\mathbf{\Gamma}_0$), and that stay within the region surrounding the mother at least at time t, so $0 < \delta\Gamma_\alpha(t) < d\Gamma$, $\forall \alpha$. The finite-time local Lyapunov exponents are:

$$\{\lambda_i[t;\mathbf{\Gamma}_0(0)]; i = 1,...2dN\} = \frac{1}{2t} \ln\left(\text{eigenvalues}\left(\mathbf{L}[t;\mathbf{\Gamma}_0(0)]^T \bullet \mathbf{L}[t;\mathbf{\Gamma}_0(0)]\right)\right), \tag{7}$$

where the notation $\lambda[t;\mathbf{\Gamma}_0(0)]$ refers to the finite-time, local Lyapunov exponent for a trajectory of length t and starting at the point $\mathbf{\Gamma}_0(0)$. If the system is chaotic, most initial points that are within an initial region will diverge from the tube at a later time (see figure 1). The escape from trajectory tubes is controlled by the sum of all the finite-time local positive Lyapunov exponents, so the probability $p(\mathbf{\Gamma}_0(0,t;d\Gamma))$ that initial phases start in the mother tube and stay within that tube is given by,

---
[1] This is true for equilibrium ensembles. The situation where this was not the case could also be considered, however a more complicated expression would result in this more general case.

$$p(\mathbf{\Gamma}_0(0,t;d\Gamma)) \propto d\Gamma^{2d_cN} f(\mathbf{\Gamma}_0,0) \exp[-\sum_{i|\lambda_i>0} \lambda_i[t;\mathbf{\Gamma}_0]t]. \tag{8}$$

Since the system is assumed to be time reversible, there will be a set of antitrajectories which are also solutions of the equations of motion. We use the notation $\mathbf{\Gamma}^*(0)$ to denote the initial phase of the antitrajectory. From figure 1 and the properties of time reversible systems we know that if $M^T\mathbf{\Gamma}_0^*(t) \equiv \mathbf{\Gamma}_0(0)$ then $M^T\mathbf{\Gamma}_0^*(0) = \mathbf{\Gamma}_0(t)$ where $M^T : M^T(\mathbf{q},\mathbf{p}) \equiv (\mathbf{q},-\mathbf{p})$ represents a time reversal mapping. Further from the time reversibility of the dynamics the set of positive Lyapunov exponents for the antitrajectory $\{\lambda_i[t;\mathbf{\Gamma}_0^*(0)]; \lambda_i > 0\}$ is identical to the set of negative exponents for the conjugate forward trajectory multiplied by $-1$,

$$\{\lambda_i[t;\mathbf{\Gamma}_0^*(0)]; \lambda_i > 0\} = \{-\lambda_j[t;\mathbf{\Gamma}_0(0)]; \lambda_j < 0\}. \tag{9}$$

Before considering the Lyapunov derivation of the TFT, it is useful to consider the computation of phase space averages of a variable A. The ensemble average, $\langle \overline{A}_t \rangle$, of the trajectory segment time average, $\overline{A}_t(\mathbf{\Gamma})$, of an arbitrary phase function $A(\mathbf{\Gamma})$, can be written as,

$$\langle \overline{A}_t \rangle = \int d\mathbf{\Gamma} \, f(\mathbf{\Gamma},0) \overline{A}_t(\mathbf{\Gamma}). \tag{10}$$

We can partition the initial phase space into $2d_CN$ - dimensional phase volume elements that are formed by the set of orthogonal eigenvectors of $\mathsf{L}[t;\mathbf{\Gamma}_0(0)]^T \bullet \mathsf{L}[t;\mathbf{\Gamma}_0(0)]$ projected from the initial mother phase points $\{\mathbf{\Gamma}_0(0)\}$. By careful construction of the partition, or mesh, we are able to ensure that each point in phase space is associated with a single mother phase - that is, it is within a region about a mother phase point $0 < \delta\Gamma_\alpha(t) < d\Gamma$, $\forall \alpha$, at least at time t. It is assumed that the phase volume elements are sufficiently small that any curvature in the direction of the eigenvectors can be ignored. In practice this phase space can be constructed as shown in figure 2. In this figure we assume that there is *no* curvature in the direction of the eigenvectors over the region considered: we expect that this will be approached as $d\Gamma \to 0$.

It should also be noted that although this diagram considers one expanding and one contracting eigendirection, there is no reason that an equal number of positive and negative exponents must exist for this construction to be used, and one or more Lyapunov exponents may be equal to zero. As long as the steady state is chaotic, the detailed structure of the steady state is irrelevant, and it is not necessary for the steady state to be Anosov.[2]

To construct the partition, an arbitrary initial mother phase point is selected and the set of points that start and remain within the tube defined by $0 < \delta\Gamma_\alpha(t) < d\Gamma, \quad \forall \alpha$ are identified. These points are considered to belong to the first region in the partition. From equation (8) it is clear that the volume occupied by these points at t = 0 is $d\Gamma^{6N} \exp[-\sum_{i|\lambda_i > 0} \lambda_i[t;\Gamma_0(0)]t]$. A second region is constructed in a similar manner, with a new mother phase point selected to be initially at a point on the corner of the first region to ensure there is no overlap of regions in the partition. Again, the set of points that start and remain within the tube defined by $0 < \delta\Gamma_\alpha(t) < d\Gamma, \quad \forall \alpha$ are identified, and a second region in the partition is constructed. This is repeated until phase space is completely partitioned into regions of volume $d\Gamma^{6N} \exp[-\sum_{i|\lambda_i > 0} \lambda_i[t;\Gamma_0(0)]t]$. Because these volumes depend on the *time-dependent, local* Lyapunov exponents, the volume of each region may differ, and the partitioning will change as longer trajectories are considered. Note that because of the continuity and uniqueness of solutions, the time evolved mesh created using this partition has the following two properties: it covers all of the initial phase space for all times, and two disjoint elements of the mesh remain disjoint for all times.[3]

Construction of the partition is shown in figure 2. In figure 2.a., a point $\Gamma_{0,1}(0)$ is selected and the region $0 < \delta\Gamma_\alpha(0) < d\Gamma, \quad \forall \alpha$ is shaded grey. The location of this region at time t is also shown. In figure 2.b., it is shown that the proportion of points that remain with in the tube emanating from $\Gamma_{0,1}(0)$ will be proportional to the Lyapunov weight, $\exp[-\sum_{\lambda_i > 0} \lambda_i[t;\Gamma_0(0)]t]$. The origin of those points are shaded in black in the region at t = 0. The black region defines the first region of the partition. In figure 2.c., a tube of equal cross-section to that in a) is formed at a new origin, $\Gamma_{0,2}(0)$, on a corner of the partition emanating from $\Gamma_{0,1}(0)$. Again the position of these points at time t is shown, and in figure 2.d., the origin of the points that remain within in the tube $0 < \delta\Gamma_\alpha(t) < d\Gamma, \quad \forall \alpha$ at time t are indicated by the hatching. The construction is repeated until phase space is covered and in figure 2.e., we show the partitioning of a small region of phase space. Note that although the phase volumes emanating from $\mathbf{\Gamma}_{0,1}(0)$ in figure 2.a. and from $\mathbf{\Gamma}_{0,2}(0)$ in figure 2.c. are square, the phase space partition associated with these points (the black region in figure 2.b and the hashed region in figure 2d respectively) are subsets of the squares. Therefore the eventual partitioning of the initial phase space is *not* a uniform grid.

---

[2] Compare this with the Chaotic Hypothesis employed by Gallavotti and Cohen [8].
[3] In contrast, the tubes of size $d\Gamma$, used to identify the regions associated with each mother phase point, will generally overlap, even at time zero, since they are of constant size but emanate from the irregularly spaced mesh of $\{\Gamma_0(0)\}$.

To calculate phase averages, it is necessary to sum over all regions, with the weight of each region given by the volume of that region (the Lyapunov weight, $\exp[-\sum_{\lambda_i>0}\lambda_i[t;\mathbf{\Gamma}_0(0)]t]$) and the initial phase space distribution function for that region. Since , for the entire range of times considered, the maximum separation of phase trajectories within each tube, from the mother trajectory is bounded by $d\Gamma$, the maximum variation of time averages and phase averages, from those of the mother trajectory, is proportional to $d\Gamma$. Thus, in the limit $d\Gamma \to 0$, we can compute $\langle \overline{A}_t \rangle$ and phase averages as,

$$\langle \overline{A}_t \rangle = \frac{\lim_{d\Gamma\to 0}\sum_{\{\mathbf{\Gamma}_0\}} \overline{A}_t(\mathbf{\Gamma}_0) f(\mathbf{\Gamma}_0(0),0)\exp[-\sum_{\lambda_i>0}\lambda_i[t;\mathbf{\Gamma}_0(0)]t]}{\lim_{d\Gamma\to 0}\sum_{\{\mathbf{\Gamma}_0\}} f(\mathbf{\Gamma}_0(0),0)\exp[-\sum_{\lambda_i>0}\lambda_i[t;\mathbf{\Gamma}_0(0)]t]} \qquad (11)$$

and

$$\langle A(s) \rangle = \frac{\lim_{d\Gamma\to 0}\sum_{\{\mathbf{\Gamma}_0(0)\}} A(\mathbf{\Gamma}_0(s)) f(\mathbf{\Gamma}_0(0),0)\exp[-\sum_{\lambda_i>0}\lambda_i[t;\mathbf{\Gamma}_0(0)]t]}{\lim_{d\Gamma\to 0}\sum_{\{\mathbf{\Gamma}_0(0)\}} f(\mathbf{\Gamma}_0(0),0)\exp[-\sum_{\lambda_i>0}\lambda_i[t;\mathbf{\Gamma}_0(0)]t]} \qquad (12)$$

respectively, where we sum over the set of mother phase points $\{\mathbf{\Gamma}_0(0)\} = \{\mathbf{\Gamma}_{0,i}(0); i=1, N_{\mathbf{\Gamma}_0}\}$. These equations simply mean that in order to obtain a phase space average, we sum over all regions in the partition, weighting each with its volume (determined from the Lyapunov weight given by equation (3) which is equivalent to the Sinai-Ruelle-Bowen (SRB) measure that is used to describe Anosov systems [1]), and multiplying by the appropriate initial distribution function.[4]

We can describe in words what the Lyapunov weights appearing in equations (11) and (12)

---

[4] Although equations (11) and (12) provide an extremely useful theoretical expression [12], due to the difficulty of constructing the partition it does not currently provide a feasible route for numerical calculation of phase averages for many particle systems.

achieve. On the set of initial phases, our partition places a greater density of initial mother phase points in those regions of greatest chaoticity - those regions with the greatest sums of positive local Lyapunov exponents. This is required because for strongly chaotic regions, trajectories diverge more quickly from the mother trajectory and we would expect time-dependent properties to vary more significantly along trajectories starting at nearby points in these regions. In order to correctly compute time averages we need to weight the time averaged properties along the mother trajectories, by the product of the initial distribution at the origin of the mother trajectory, and the measure of the initial hypervolume of those trajectories which do not escape from the mother trajectory. These volumes are proportional to the *negative* exponentials of the sums of positive local Lyapunov exponents. This mesh will ensure that the variation of time averages of a property along trajectories within the same mesh element will vary on a similar order to averages of the property within a partition element at the initial time [12].

We now apply these concepts to compute the ratio of conjugate averages of the dissipation function. The ratio of the probabilities of observing the two volume elements about $\mathbf{\Gamma}(0)$ and $\mathbf{\Gamma}^*(0) = M^T(\mathbf{\Gamma}(t))$ at time zero is:

$$\frac{p(\delta V_\Gamma(\mathbf{\Gamma}(0),0))}{p(\delta V_\Gamma(\mathbf{\Gamma}^*(0),0))} = \frac{f(\mathbf{\Gamma}(0),0)\delta V_\Gamma(\mathbf{\Gamma}(0),0)}{f(\mathbf{\Gamma}^*(0),0)\delta V_\Gamma(\mathbf{\Gamma}^*(0),0)}. \tag{13}$$

In deriving (13) [5], we assume:

- The initial distribution $f(\mathbf{\Gamma},0)$ is symmetric under the time reversal mapping ($f(\mathbf{\Gamma},0) = f(M^T(\mathbf{\Gamma}),0)$)[5] [Note: The initial phase space distribution does *not* have to be an equilibrium distribution.];

- The equations of motion (1), are reversible;[6] and,

- The initial ensemble and the subsequent dynamics are *ergodically consistent* [9]:

---

[5] If this is not the case, a more general form of equation (8) and hence the FT (12) can still be obtained. Equation (13) becomes $\frac{p[\delta V_\Gamma(\mathbf{\Gamma}(0),0]}{p[\delta V_\Gamma(\mathbf{\Gamma}^*(0),0]} = \frac{f[\mathbf{\Gamma}(0),0]\delta V_\Gamma(\mathbf{\Gamma}(0),0)}{f[M^T(\mathbf{\Gamma}(t)),0]\delta V_\Gamma(\mathbf{\Gamma}^*(0),0)}$. Furthermore, alternative reversal mappings to the time reversal map $M^T$ (such as the Kawasaki map [11; 3]) may be necessary to generate the conjugate trajectories in some situations [3, 13].

[6] Note that a looser condition, that will still lead to (2.1.5) and (2.1.7), is that the reverse trajectory *must exist*. This enables the proof to be extended to stochastic dynamics [4a, 14].

$$f(M^T[\mathbf{\Gamma}(t)],0) = f(\mathbf{\Gamma}^*(0),0) \neq 0, \forall \mathbf{\Gamma}(0). \tag{14}$$

Ergodic consistency (14) requires that the initial ensemble must actually <u>contain</u> time reversed phases of all possible trajectory end points. Ergodic consistency would be violated for example, if the initial ensemble was microcanonical but the subsequent dynamics was adiabatic and therefore did not preserve the energy of the system.[7]

It is convenient to define a dissipation function $\Omega(\mathbf{\Gamma})$,

$$\int_0^t ds\, \Omega(\mathbf{\Gamma}(s)) \equiv \ln\left(\frac{f(\mathbf{\Gamma}(0),0)}{f(\mathbf{\Gamma}(t),0)}\right) - \int_0^t ds\, \Lambda(\mathbf{\Gamma}(s))$$
$$= \overline{\Omega}_t t \tag{15}$$

where $\Lambda(\mathbf{\Gamma})$ is the phase space compression factor at $\mathbf{\Gamma}$. We can now calculate the probability ratio for observing a particular time averaged value A, of the dissipation function $\overline{\Omega}_t$ and its negative, $-A$, to within some fine tolerance, dA. This is achieved by dividing the initial phase space into subregions $\{\delta V_\Gamma(\mathbf{\Gamma}_i); i = 1,..\}$ centred on an initial set of phases $\{\mathbf{\Gamma}_{i,0}(0); i = 1,..\} \equiv \{\mathbf{\Gamma}_0(0)\}$. The probability ratio can be obtained by calculating the corresponding ratio of probabilities that the system is found <u>initially</u> in those subregions which <u>subsequently</u> generate bundles of trajectory segments with the requisite time average values of the dissipation function. Thus the probability of observing the complementary time average values of the dissipation function is given by the ratio of generating the <u>initial</u> phases from which the <u>subsequent</u> trajectories evolve. We now sum over all subregions for which the time averaged dissipation function takes on the specified values,

---

[7] Jarzynski [15] and Crooks [16] treat cases where the dynamics is not ergodically consistent and thereby obtain expressions for Helmholtz free energy differences between *different* systems [17]. This work has been widely applied and extended, see for example [18].

$$\frac{\Pr(\overline{\Omega}_t = A)}{\Pr(\overline{\Omega}_t = -A)} = \lim_{d\Gamma \to 0} \frac{\sum_{\{\Gamma_0 | \overline{\Omega}_t(\Gamma_0) = A\}} f(\Gamma_0(0),0) \exp[-\sum_{\lambda_j > 0} \lambda_j[t;\Gamma_0(0)]t]}{\sum_{\{\Gamma_0 | \overline{\Omega}_t(\Gamma_0) = -A\}} f(\Gamma_0(0),0) \exp[-\sum_{\lambda_j > 0} \lambda_j[t;\Gamma_0(0)]t]}$$

$$= \lim_{d\Gamma \to 0} \frac{\sum_{\{\Gamma_0 | \overline{\Omega}_t(\Gamma_0) = A\}} f(\Gamma_0(0),0) \exp[-\sum_{\lambda_j > 0} \lambda_j[t;\Gamma_0(0)]t]}{\sum_{\{\Gamma_0 | \overline{\Omega}_t(\Gamma_0) = A\}} f(\Gamma_0^*(0),0) \exp[-\sum_{\lambda_j > 0} \lambda_j[t;\Gamma_0^*(0)]t]} \quad (16)$$

$$= \lim_{d\Gamma \to 0} \frac{\sum_{\{\Gamma_0 | \overline{\Omega}_t(\Gamma_0) = A\}} f(\Gamma_0(0),0) \exp[-\sum_{\lambda_j > 0} \lambda_j[t;\Gamma_0(0)]t]}{\sum_{\{\Gamma_0 | \overline{\Omega}_t(\Gamma_0) = A\}} f(\Gamma_0(t),0) \exp[+\sum_{\lambda_j < 0} \lambda_j[t;\Gamma_0(0)]t]}$$

where we use the relationships between conjugate trajectories to express the numerator and denominator in terms of sums over $\{\Gamma_0 | \overline{\Omega}_t(\Gamma_0) = A\}$. The notation $\sum_{\{\Gamma_0 | \overline{\Omega}_t = A\}} \ldots$ is used to indicate that the sum is carried out over the set of regions in the mesh for which $\overline{\Omega}_t = A \pm dA$. Because of the method of construction of our partition, the tolerance $dA$ is proportional to $d\Gamma$.

In the first line of (16) there is of course no guarantee that initial phases $\Gamma_0(0), \Gamma_0^*(0) \equiv M^T \Gamma_0(t)$ with conjugate values for time dissipation function $\pm A$, will *both* be vertex mesh points on our partition of the initial phase space - see Figure 2. However, in the limit $d\Gamma \to 0$, this requirement can be satisfied arbitrarily closely. If $\Gamma_0(0)$ is a vertex mesh point for which $\overline{\Omega}_t(\Gamma_0) = A$, then in the limit $d\Gamma \to 0$, there will be another vertex mesh point which is arbitrarily close to $\Gamma_0^*(0)$ and within whose cell, time average values of the dissipation function will be, within a fine tolerance $dA \propto d\Gamma$, equal to $-A \pm dA$.

Using (15) to substitute for $f(\Gamma_0(t),0)$, we obtain,

$$\frac{\Pr(\overline{\Omega}_t = A)}{\Pr(\overline{\Omega}_t = -A)} = \lim_{d\Gamma \to 0} \frac{\sum_{\{\Gamma_0 | \overline{\Omega}_t(\Gamma_0) = A\}} f(\Gamma_0(0),0) \exp[-\sum_{\lambda_i > 0} \lambda_i[t;\Gamma_0(0)]t]}{\sum_{\{\Gamma_0 | \overline{\Omega}_t(\Gamma_0) = A\}} \exp[-\overline{\Omega}_t t] f(\Gamma_0(0),0) \exp[-\overline{\Lambda}_t t] \exp[+\sum_{\lambda_i < 0} \lambda_i[t;\Gamma_0(0)]t]}$$

$$= \lim_{d\Gamma \to 0} \exp[At] \frac{\sum_{\{\Gamma_0 | \overline{\Omega}_t(\Gamma_0) = A\}} f(\Gamma_0(0),0) \exp[-\sum_{\lambda_i > 0} \lambda_i[t;\Gamma_0(0)]t]}{\sum_{\{\Gamma_0 | \overline{\Omega}_t(\Gamma_0) = A\}} f(\Gamma_0(0),0) \exp[-\sum_{\lambda_i > 0} \lambda_i[t;\Gamma_0(0)]t]} \qquad (17)$$

$$= \exp[At]$$

To obtain the second line we use the fact that the sum of <u>all</u> the local Lyapunov exponents is the time average of the phase space compression factor: $\overline{\Lambda}_t(\Gamma_0) = \sum_{\forall i} \lambda_i[t;\Gamma_0]$.

Of course (17) is identical to the ensemble independent TFT derived previously [5, 9]. For an isoenergetic system, the SSFT derived from (17) is identical to that obtained previously for this system [1, 8] (see below). However, if the initial ensemble is not microcanonical, the Lyapunov weights and associated SRB measure do <u>not</u> dominate the weight that results from the nonuniformity of the initial distribution. This is evidenced by the fact that when the initial ensemble is *not* microcanonical, $\Omega(\Gamma) \neq \Lambda(\Gamma)$. For ensembles other than the microcanonical ensemble the Fluctuation Theorem given in (17) has been confirmed numerically whereas the corresponding "theorems" employing the phase space compression factor have not, [5].

## 3     THE STEADY STATE FT AND ERGODICITY

Previous Lyapunov derivations of SSFT [1, 8], assumed either that the initial probability distribution was uniform (e.g. microcanonical), or if nonuniform, that variations in the initial density could be ignored at long times and the probability of observing trajectory segments would always be dominated by the exponential of the sum of positive Lyapunov exponents (*i.e.* it was assumed that the positive exponential escape from phase space trajectory tubes would always dominate in the long time limit). Here we show that this is not the case and that consistent with the Liouville derivation of the FT [5], the steady state FT does indeed depend on the initial ensemble and the dynamics of the system. For an isoenergetic system, the results obtained are identical to those obtained previously for this system [1-3, 5, 8].

We note that in the TFT, time averages are carried out from $t = 0$ (where we have an initial distribution $f(\Gamma,0)$), to some arbitrary later time t - see (15). One can make the averaging time arbitrarily long. For sufficiently long averaging times t, we might *approximate* the time averages in (15) by performing the time average not from $t = 0$ but from some later time $\tau_R << t$,

$$\overline{\Omega}_{t-\tau_R}(\tau_R) \equiv \frac{1}{t-\tau_R} \int_{\tau_R}^{t} ds\, \Omega(s). \tag{18}$$

where the notation $\overline{\Omega}_{t-\tau_R}(\tau_R)$ implies that the averaging time is $t - \tau_R$, and commences at time $\tau_R$. Thus we see that,

$$\begin{aligned}\overline{\Omega}_t &\equiv \frac{1}{t}\int_0^t ds\, \Omega(s) \\ &= \frac{1}{t-\tau_R}\int_{\tau_R}^t ds\, \Omega(s) + O(\tau_R/t) \\ &= \overline{\Omega}_{t-\tau_R}(\tau_R) + O(\tau_R/t)\end{aligned} \tag{19}$$

Now for systems with finite correlation times (usually $O(\tau_R)$), the distribution of values of $\overline{\Omega}_t$ will in the long time limit, be approximately normally distributed about $\overline{\Omega}_\infty$ with a standard deviation that is proportional to $t^{-1/2}$. For such systems (17, 19), can be combined to give an asymptotic form of the FT,

$$\lim_{t/\tau_R \to \infty} \frac{1}{t} \ln \frac{p(\overline{\Omega}_{t-\tau_R}(\tau_R) = A)}{p(\overline{\Omega}_{t-\tau_R}(\tau_R) = -A)} = A. \tag{20}$$

In the long time limit the error in making the approximation in (19) is asymptotically insignificant compared to the range of observeable values for $\overline{\Omega}_t$.

If the system is thermostatted in some way and if after some finite transient relaxation time $\tau_R$ (which is short compared to t), it comes to a nonequilibrium *steady state*, then (20) is in fact an asymptotic Steady State Fluctuation Theorem (SSFT),

$$\lim_{t/\tau_R \to \infty} \frac{1}{t} \ln \frac{p(\overline{\Omega}_{t,ss} = A)}{p(\overline{\Omega}_{t,ss} = -A)} = A. \tag{21}$$

In this equation $\overline{\Omega}_{t,ss}$ denotes the fact that the time averages are only computed after the relaxation of initial transients (*i.e.* in a nonequilibrium steady state). It is understood that the probabilities are computed over an *ensemble* of long trajectories which initially (at some long time in the past) were characterised by the distribution $f(\Gamma,0)$ at t = 0. A more rigourous derivation of (21) from (17) can be obtained in some cases [19].

We often expect that the nonequilibrium steady state is unique or *ergodic*. When this is so, steady state time averages and statistics are independent of (almost) any initial starting phase at t = 0. Most of nonequilibrium statistical mechanics is based on the assumption that the systems being studied are *ergodic*. For example the Chapman Enskog solution of the Boltzmann equation is based on the tacit assumption of ergodicity. Experimentally, one does not usually measure transport coefficients as ensemble averages; almost universally transport coefficients are measured as time averages, although experimentalists often employ repeated experiments under identical macroscopic conditions in order to determine the statistical uncertainties in their measured time averages. They would not expect that the results of their measurements would depend on the initial (unspecifiable!) microstate. Arguably, the clearest indication of the ubiquity of nonequilibrium ergodicity, is that empirical data tabulations assume that transport coefficients are single valued functions of the macrostate: (N,V,T) and possibly the strength of the dissipative field. The tacit assumption of nonequilibrium ergodicity is so widespread that it is frequently forgotten that it is in fact an *assumption*. The necessary and sufficient conditions for ergodicity are not known. However if the initial ensemble used to obtain equation (21) is the equilibrium ensemble generated by the dynamics when the nonequilibrium driving force is removed,[8] and the system is ergodic then the probabilities referred to in the SSFT (21) can be computed not only over an *ensemble* of trajectories, but also over segments along a single exceedingly long phase space trajectory (that can be divided into segments, each of which has length t $\gg \tau_M$).

## 4. CONCLUSIONS

We have given a derivation of a Transient Fluctuation Theorem using Lyapunov weights. In addition we have shown that by taking the long time limit, a FT for an ensemble of steady states trajectories can be obtained, and have argued that if the steady state is ergodic and the appropriate initial ensemble is used in the derivation, a Steady State Fluctuation Theorem that applies to trajectory segments along a single trajectory can be obtained. The theorems we so derive are identical to those derived by Evans and Searles previously using Gibbs weights [2, 3, 5, 9]. They apply to a wide variety of initial ensembles and dynamics (*e.g.* constant energy or constant temperature).

In contrast to the derivation of Gallavotti and Cohen (GC) [8] which treats the nonequilibrium steady state attractor, the present derivation assumes a known smooth initial phase space distribution. This simplifies the derivation somewhat, because the steady state attractor treated in the GC derivation is known to be fractal. However, the price we pay for this simplification is that in both our TFT and the SSFT, we deal with the fluctuation properties of an *ensemble* of trajectories rather than with a single (very long) dynamical trajectory as in GC. We argue that if the systems studied are ergodic, then our ensemble-based SSFT (21), should be applicable to the case of a single very long steady state dynamical trajectory.

Both the transient and steady state ensemble-based forms of the FT have been numerically verified for systems under various combinations of ensemble and dynamics [5, 7]. In the case of steady states, numerical verification has been successful for both the ensemble version of the SSFT (21) and for the case of a single dynamical trajectory. Furthermore, the TFT has recently be verified experimentally using optical tweezers apparatus [10]. We therefore have considerable confidence in the correctness of all the ensemble-based FT's that we have derived, and their applicability to single dynamical steady state trajectories under a variety of thermostats and barostats etc.

---

[8] This requires more than ergodic consistency (see equation (14)) that is required to generate the TFT and the ensemble version of the SSFT. It means for example, if the steady state is isoenergetic, then the microcanonical ensemble must be used as the initial ensemble - a canonical initial distribution is ergodically consistent with isoenergetic dynamics, but would not be suitable for generation of the *dynamic* version of the SSFT because it would generate a set of isoenergetic steady states with different energies. This condition can be expressed by stating that there is a unique steady state for the selected combination of initial ensemble and dynamics.

There is however, at least one unresolved issue. Provided the system is sufficiently Anosov-like, the GC derivation of the SSFT is independent of the form of the dynamics. The GC proof applies equally well to isoenergetic or to isothermal dynamics. However, the GC statement of the SSFT says that provided the system obeys the *chaotic hypothesis*, then

$$\lim_{t \to \infty} \frac{1}{t} \ln \frac{p(\overline{\Lambda}_{t,ss} = A)}{p(\overline{\Lambda}_{t,ss} = -A)} = A. \tag{22}$$

However (22) has only been verified numerically for isoenergetic dynamics. Furthermore for both an ensemble of steady state trajectories and for a single dynamical trajectory, there is not yet any convincing simulation evidence that (22) is correct for non-isoenergetic (*e.g.* isokinetic) dynamics [5]. It is clear from numerical studies that at small fields, if (22) is correct, the convergence is very slow compared to that of equation (21). The resolution of this issue is currently under intensive investigation.

**Acknowledgements**

We wish to acknowledge L. Rondoni for useful comments. We also thank the Australian Research Council.

**Figure Captions**

**Figure 1** A schematic diagram showing how a trajectory and its conjugate evolve. The square region emanating from $\mathbf{\Gamma}_0(0)$ has axes aligned with the eigenvectors of the tangent vector propagator matrix $\mathbf{L}[t;\mathbf{\Gamma}_0(0)]^T \bullet \mathbf{L}[t;\mathbf{\Gamma}_0(0)]$. The shaded region thus shows where initial points in this region will propagate to at time t. For illustrative purposes we assume a two dimensional ostensible phase space and that there is one positive time-dependent local Lyapunov exponent (in the x-direction for the trajectory starting at $\mathbf{\Gamma}_0$) and one negative time-dependent local Lyapunov exponent (in the y-direction for the trajectory starting at $\mathbf{\Gamma}_0$).

**Figure 2** A schematic diagram showing the construction of the partition, or mesh, used to determine phase space averages using Lyapunov weights. For convenience, we assume a two dimensional phase space, and that there is one positive and one negative time-dependent local Lyapunov exponent for each region in the section of phase space shown. In a) and b) the construction of the first partition is shown. This partition (shown in black in b)) emanates from $\mathbf{\Gamma}_{0,1}(0)$ and contains all points that remain in the region $0 < \delta\Gamma_\alpha(0) < d\Gamma, \quad \forall \alpha$. In c) and d) construction of the second partition (shown as a hatched region) is shown and in e) the partitioning of phase space after several iterations of this process is shown. The size of phase volume elements is assumed to be sufficiently small that any curvature in the direction of the eigenvectors can be ignored.

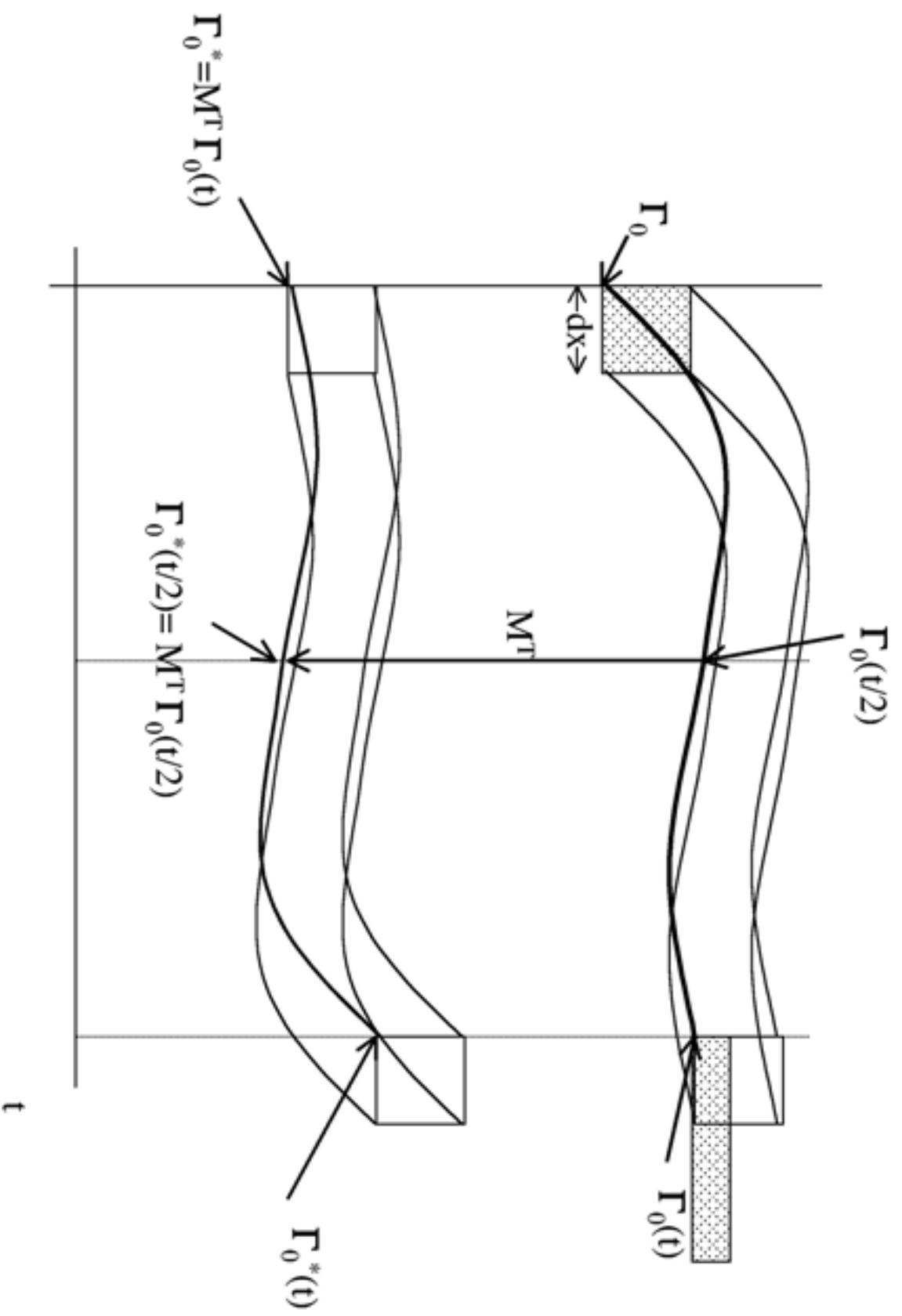

Figure 1 Jepps, Evans and Searles

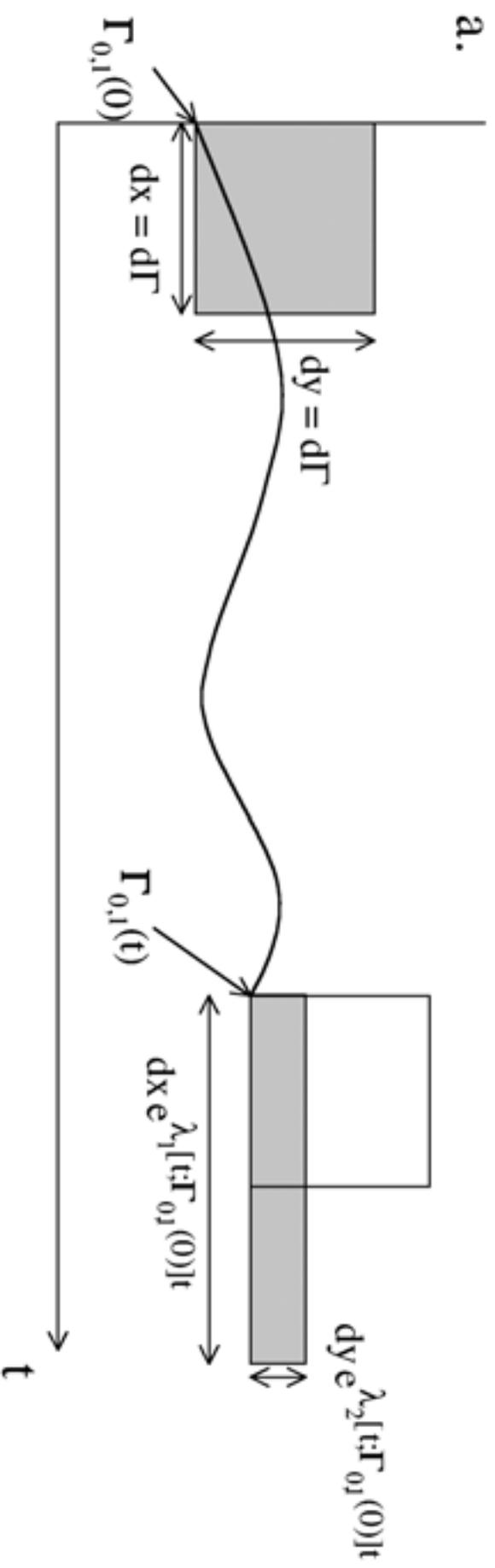

Figure 2a Jepps, Evans and Searles

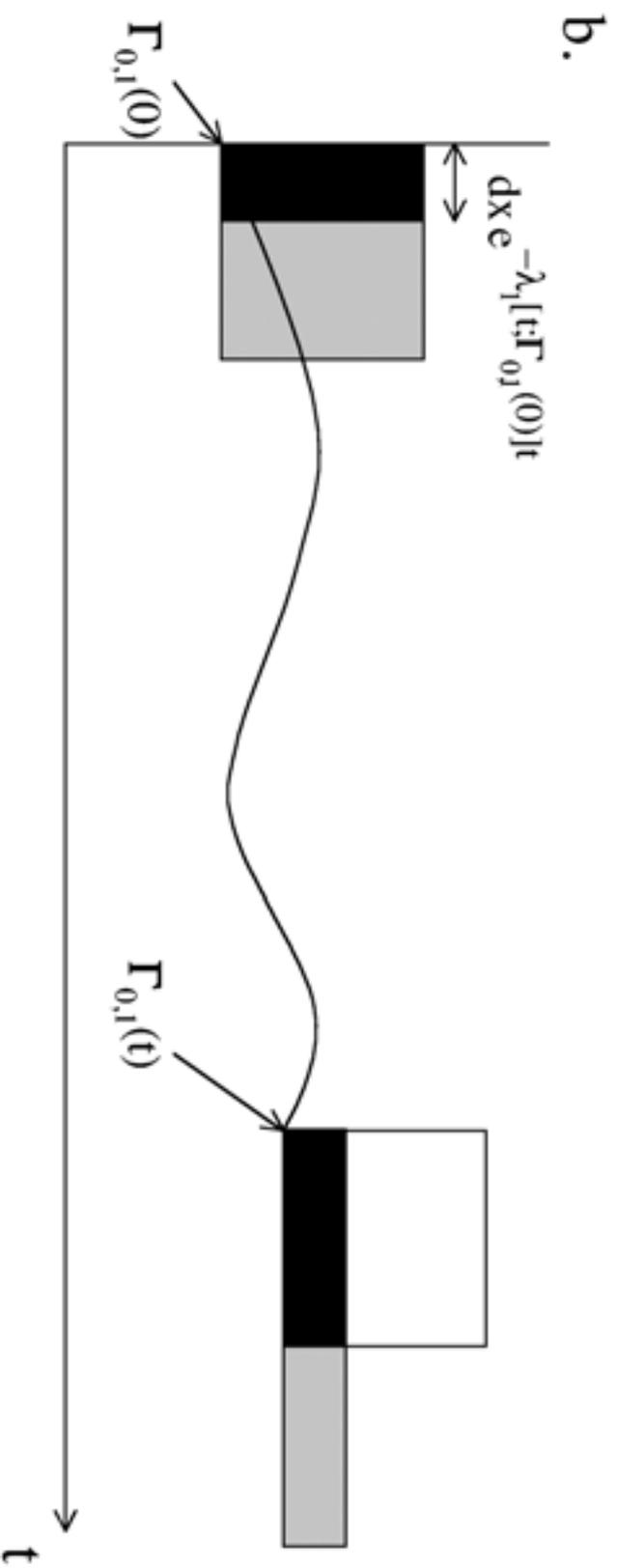

Figure 2b Jepps, Evans and Searles

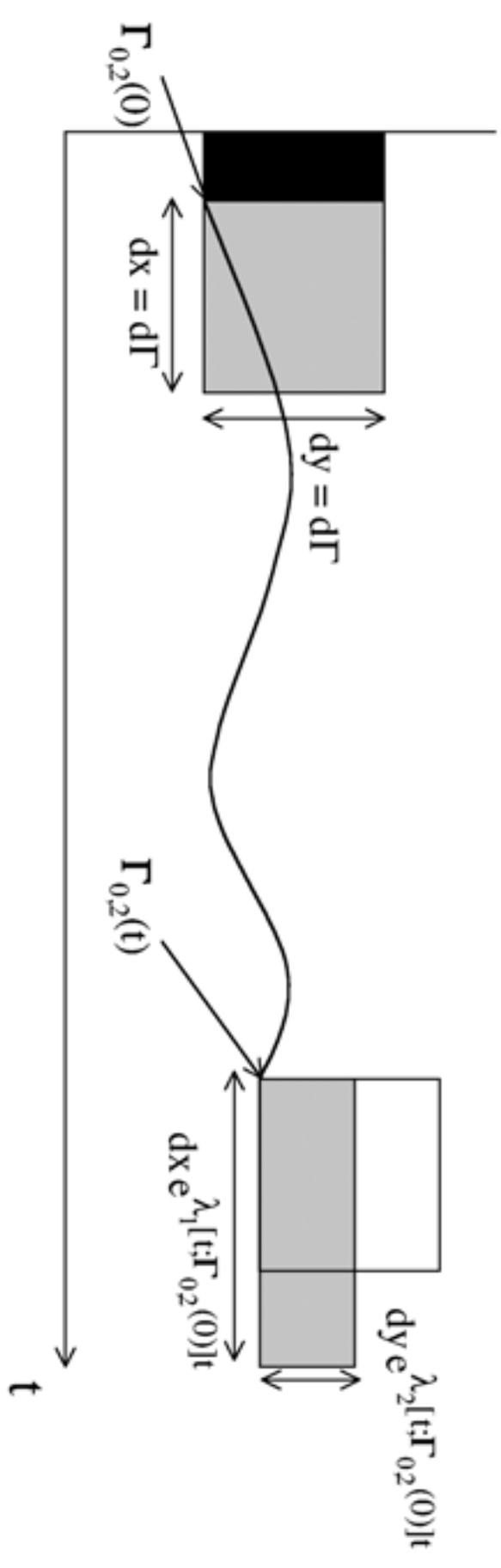

Figure 2c Jepps, Evans and Searles

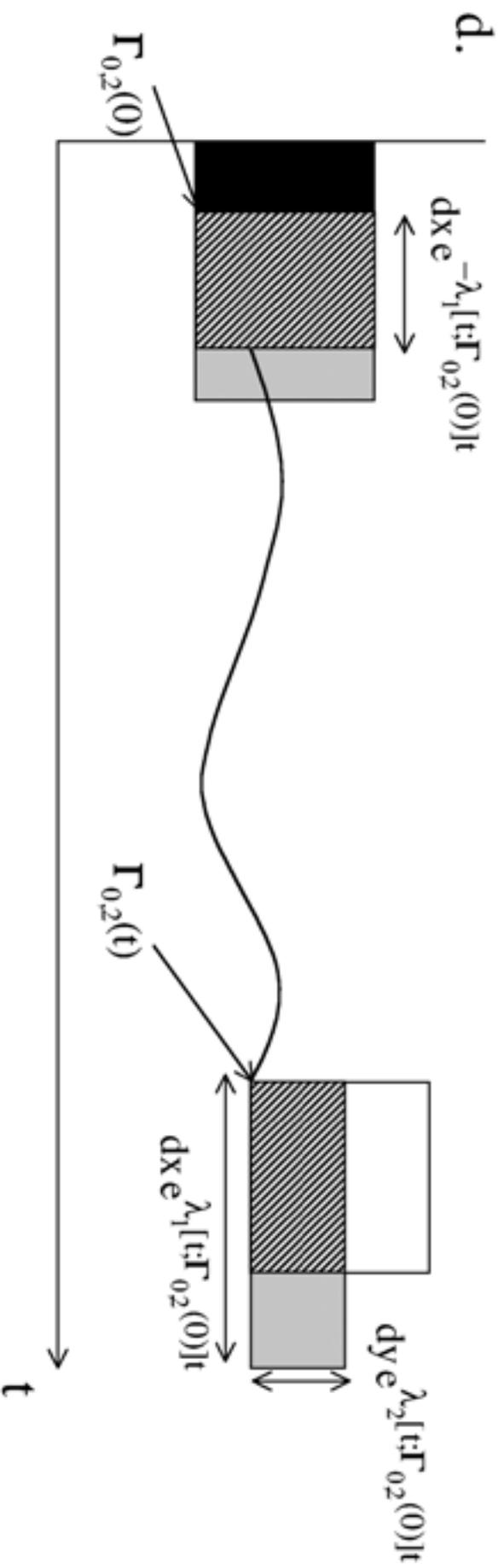

Figure 2d Jepps, Evans and Searles

e.

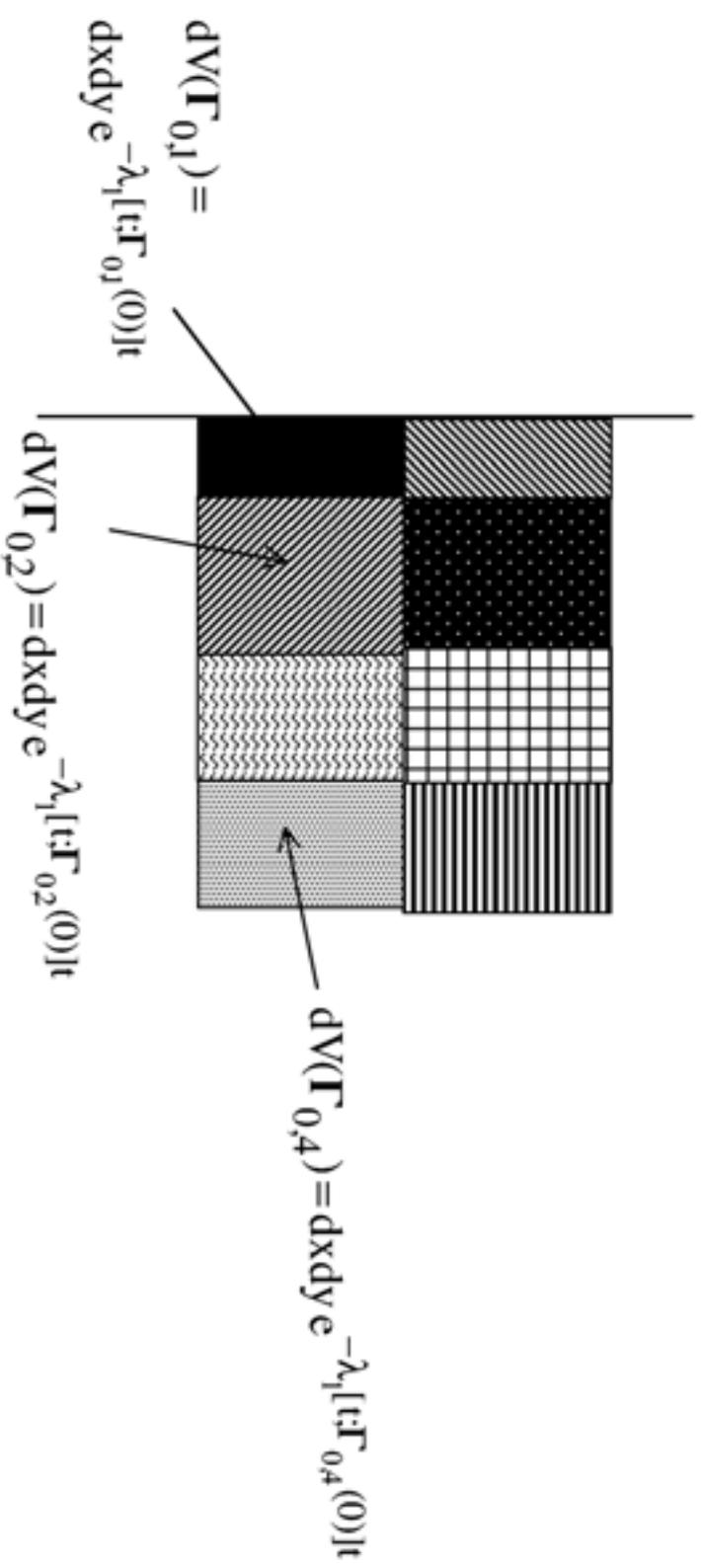

$dV(\Gamma_{0,1}) =$
$dxdy e^{-\lambda_1[t;\Gamma_{0,1}(0)]t}$

$dV(\Gamma_{0,2}) = dxdy e^{-\lambda_1[t;\Gamma_{0,2}(0)]t}$

$dV(\Gamma_{0,4}) = dxdy e^{-\lambda_1[t;\Gamma_{0,4}(0)]t}$

Figure 2e Jepps, Evans and Searles